\documentclass[prl,twocolumn,showpacs,preprintnumbers,amsmath,amssymb,superscriptaddress,footinbib]{revtex4-1}

\usepackage{graphicx}
\usepackage{dcolumn}
\usepackage{bm}
\usepackage{amsmath}
\usepackage[usenames]{color}

\begin{document}

\title{Scale-Free Structure Emerging from Co-Evolution of a Network and the Distribution of a Diffusive Resource on It}
\author{Takaaki Aoki}
\email{takaaki.aoki.work@gmail.com}
\affiliation{Faculty of Education, Kagawa University, Takamatsu 760-8521, Japan}
\author{Toshio Aoyagi}
\affiliation{Graduate School of Informatics, Kyoto University, Kyoto 606-8501, Japan}
\affiliation{CREST, Japan Science and Technology Agency, Kawaguchi, Saitama 332-0012, Japan}

\date{2011-07-06}
\begin{abstract}
Co-evolution exhibited by a network system, involving the intricate interplay between the dynamics of the network itself and the subsystems connected by it, is a key concept for understanding the self-organized, flexible nature of real-world network systems. We propose a simple model of such co-evolving network dynamics, in which the diffusion of a resource over a weighted network and the resource-driven evolution of the link weights occur simultaneously.  We demonstrate that, under feasible conditions,  the network robustly acquires scale-free characteristics in the asymptotic state. Interestingly, in the case that the system includes dissipation, it asymptotically realizes a dynamical phase characterized by an organized scale-free network, in which the ranking of each node with respect to the quantity of the resource possessed thereby changes ceaselessly. Our model offers a unified framework for understanding some real-world diffusion-driven network systems of diverse types.
\end{abstract}

\maketitle

Today, the term ``network'' is common in our everyday lives, in which it often refers to large-scale, complexly structured, conglomerations of interactions in real-world systems.
Typically, these networks are not static, but change continuously in response to the activity of the subsystems that they connect.
For example,  traffic networks among cities supporting the transportation of people and products
 are frequently reformed to meet the current needs as cities develop or decay,
and conversely, this reformation of the networks influences the growth or decay of the cities.
A similar process takes place in the case of communication networks \cite{Kalapala:2006ie,*Barrat:2004bk}.
In social networks, human behavior is strongly influenced by social relationships,
and at the same time, the relationships among people change continually as a result of their behavior.
This intricate interplay between individuals and their relationships creates the complex structures of human societies \cite{Palla:2007if,*Kossinets:2006je}.
The essence of such real-world systems resides in the co-evolving dynamics of the individual subsystems
and the networks of interactions through which they are connected.
To understand the mechanisms governing such dynamical network organization,
we have to consider the interplay between the dynamics both on and of a network.

In the last decade, there have been two major trends in the investigation of the type of co-evolving dynamics described above.
One trend is to focus on the topology of the network.
It is well known that  real-world networks possess some common topological features \cite{Watts:1998db,Barabasi:1999p178,Newman:2003p839}.
One such feature is a scale-free structure, in which the ``node degree'' (the total number of links connected to a node)
exhibits a power-law distribution \cite{PRICE:1965vs,Barabasi:1999p178}.
In the attempt to explain these features, many models describing the evolution of the topology have been proposed.
The other trend is to focus on the collective behavior of the dynamical subsystems interacting on complex but static networks \cite{Boccaletti:2006p403}.
The aim of such studies is to investigate how various observed topological features
of the network influence the nature of the dynamical systems coupled through the links of the network.

Most studies employing the approaches described above focus on only one of the two aspects of co-evolving dynamics, evolution of the network topology or evolution of the dynamical states of the nodes on a static network.
However, co-evolving dynamics, in which the network topology and the nodal states evolve simultaneously
and interdependently, is an interdisciplinary subject of growing interest \cite{Gross:2008p274,*Bohme:2011gw,*Vazquez:2008dh,*Nardini:2008jn,*Holme:2006wi,*Aoki:2009p4,*Perc:2010p822,*PhysRevE.84.036101}.
To facilitate further systematic studies of such systems, a general mathematical framework for  modeling  co-evolving real-world networks is needed.
As a first step toward this goal, in this Letter,
we propose a simple model of co-evolving weighted networks. This model is schematically depicted in Fig.1(a).
The basic concept of our model is as follows. 
We assign a dynamical variable, $x_i$, to each node. The dynamics of these variables are governed by a reaction-diffusion equation in which nodes are coupled through the weighted links of the network.
This dynamical variable at each node can be regarded as the quantity of the ``resource'' at that node.
Additionally, we assume a physically reasonable resource-dependent dynamics for the link weights.
We systematically investigate the collective behavior that emerges asymptotically through the interplay between these dynamics both on and of the network.

\begin{figure}[t]
\includegraphics{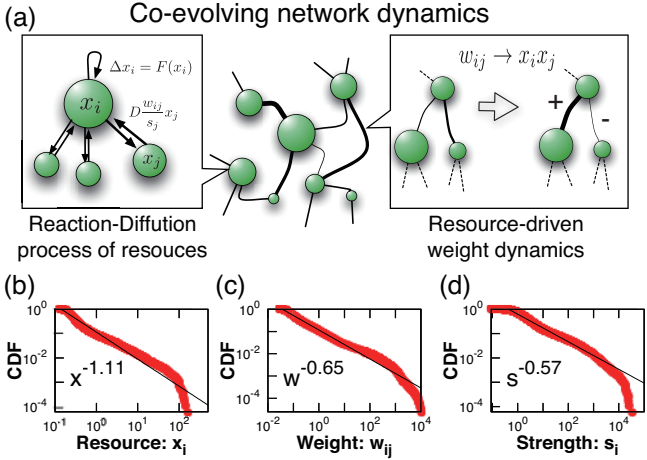}
\caption{(color online).
(a)
A schematic illustration of co-evolving dynamics of a network and the distribution of a diffusive resource on it.
(b, c, d)
Typical features of the network that emerge through the co-evolving dynamics.
The cumulative distributions of the quantity of the resource, $x_{i}$, the weights, $w_{ij}$, and the strengths, $s_{i}$, converge to power-law forms.
The topology of the network, represented by  the adjacency matrix $a_{ij}$, is chosen as an Erd\"{o}s-R\'{e}nyi random graph with $N$ = 16384 and $\langle k \rangle$ = 10.
The initial distribution of the resource and the initial weights were generated
according to a normal distribution with mean $\mu$ = 1 and standard deviation $\sigma$ = 0.1.
Other parameter values are as follows: $\kappa$ = 0.05, $D$ = 0.34, $\epsilon$ = 0.01.
}
\label{fig:CoEv}
\end{figure}

We now describe the model.
First, let us consider a weighted network with $N$ nodes. 
The link structure of the network is defined by the adjacency matrix $a_{ij}$,
in which $a_{ij}$=1 if a link exists between the nodes $i$ and $j$ and $a_{ij}$=0 otherwise.
We assign a time-dependent symmetric weight $w_{ij}(t)$ ($=w_{ji}(t))$ to each existing link.
This weight represents the strength of the interaction.
Here we study a system in which it is these strengths of the interactions that change in time.
More precisely, we consider a system in which there exists a fixed set of connections among nodes, with each connection characterized by a weight, $w_{ij}(t)$. 

We consider the reaction-diffusion dynamics of a single quantity on this network.
We refer to this quantity as the ``resource",
which may be, for example,  molecules, cells, people or money.
The value of this quantity at the $i$th node at time $t$ is represented by $x_i(t)$.
In general, the evolution of $x_i(t)$ is assumed to be described by an equation of the following form:
$$
\Delta x_i (t)= F(x_i(t) )+ \text{diffusion process via weighted links},
$$
where $\Delta x_i(t) \equiv x_i(t+1) - x_i(t)$. Here $F(x)$ represents a reaction process undergone by the resource.
The weights $w_{ij}(t)$ control the diffusion process as follows.
This process  can be understood as consisting of the combined motion of many random walkers,
in which the walkers at the node $i$ move to the node $j$ in a single time step with the time-dependent probability $D w_{ji}(t) / s_i(t)$.
Here, $s_i(t)$ is the strength of the node $i$, defined by $s_i (t)\equiv \sum_{j \in {\cal N}_{i}} w_{ji}(t) = \sum_j a_{ij}w_{ij}(t)$,  where ${\cal N}_{i}$ is the set of nodes connected to the node $i$.
The master equation for the resource is thus given by
\begin{align}
   \Delta x_{i}(t)  = F(x_{i}(t) ) + 	 D \sum_{j \in {\mathcal N}_{i} } \left(  \frac{w_{ij}(t)}{s_{j}(t)} x_j(t)  - \frac{w_{ji}(t)}{s_{i}(t)}  x_i(t) \right),
\end{align}
where the second and third terms are the inward and outward currents of the resource at the $i$-th node, respectively. For the reaction process,  we employ simple dissipation with equilibrium state $x=1$, described by 
$$F(x) = -\kappa(x -1).$$

Next, we describe the evolution of the structure of the network.
It is reasonable to assume that the evolution of a weight $w_{ij}(t)$ depends on the quantities of the resource at the corresponding nodes, $x_i(t)$ and $x_j(t)$.
As a first step, we assume a linear dependence on each, with
the weight $w_{ij}(t)$ merely relaxing to $x_i(t) x_j(t)$, appealing to the law of mass action.
Then, employing the simplest form, we stipulate the dynamics of the weights to be described by
$$
w_{ij}(t+1)  - w_{ij}(t) =  \epsilon \left[ x_i(t) x_j(t) - w_{ij}(t) \right], \eqno(2)
$$
where the parameter $\epsilon^{-1}$ represents the relaxation time scale of the weight dynamics.
It should be noted that in these dynamics it is possible for a link to be effectively eliminated,
because weights can become vanishingly small.

The co-evolving dynamics of the entire system are described by the simple equations (1) and (2).
Despite their simplicity, however, we have found that the interplay between the two types of dynamics
that they describe can yield power-law distributions of the resource and the weights,
even when the underlying topology of the network is not scale-free.
Figures \ref{fig:CoEv}(b)-(d) display a typical result of the numerical simulations,
where an Erd\"{o}s-R\'{e}nyi (ER) random graph was used for $a_{ij}$.
As displayed in Fig. \ref{fig:CoEv}(b), the cumulative distribution of the resource takes a power-law form with exponent $\gamma \sim -1$ in the asymptotic state.
This result is consistent with an empirical law found to characterize many physical and
social phenomena, including word frequencies in natural languages, populations of
cities, statistics of Web access, and company sizes \cite{Miller:1958p1240,*Gabaix:1999p1317,*Huberman:1998p1328,*Aoyama:2000p1361,*Dragulescu:2001p1511,*PhysRevLett.108.168701}, namely, Zipf's law \cite{zipforig}.
As seen in Figs. \ref{fig:CoEv}(c) and (d), the weights $w_{ij}$ and strengths $s_{i}$ also exhibit power-law distributions in the asymptotic state with different exponents.
For details of the network dynamics, see Figure S1 and the movie included in the Supplemental Material.

\begin{figure*}[t]
\includegraphics{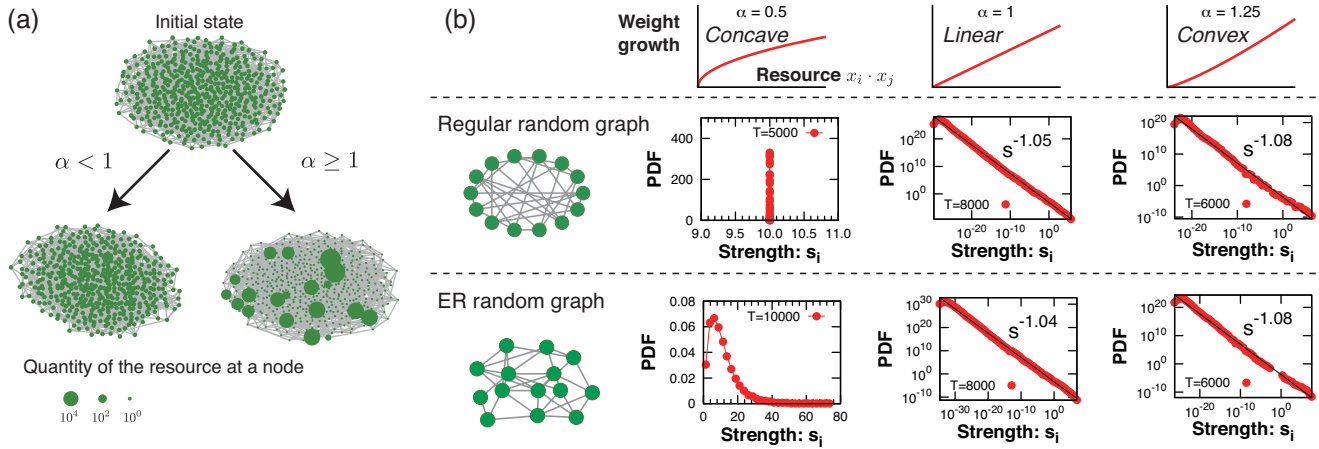}
\caption{(color online).
In the extreme case $\kappa = 0$, in which the resource diffuses over the network without dissipation,
we find two regimes, corresponding to different types of dependence of $w_{ij}(t)$ on $x_i(t)$ and $x_j(t)$,
characterized by the value of a parameter $\alpha$ (see Eq. (3)).
In these two regimes, qualitatively different network structures are realized in the asymptotic state.
(a) The co-evolving
network dynamics yields a scale-free structure with a power-law resource distribution for $\alpha$ = 1.25 and a
non-scale-free structure for $\alpha$ = 0.5. The underlying topology, $a_{ij}$, is given by a regular random graph ($N$ = 512, $k$ = 5).
(b) The dependence of the asymptotic strength distribution on the imposed topology of the network and the type of the weight dynamics.
The top three graphs correspond to three types of weight dependence  of $\Delta w_{ij}$ on $x_i$ and $x_j$:
$\Delta w_{ij} \sim (x_{i} x_{j})^{\alpha} $
with $\alpha$ = 0.75, 1 and 1.25.
The other parameter values are as follows: $N$ = 16384, $\langle k \rangle$ =10, $D$ = 0.02, $\epsilon$ = 0.01. }
\label{fig:DiffusionOnly}
\end{figure*}

\begin{figure}[t]
\includegraphics{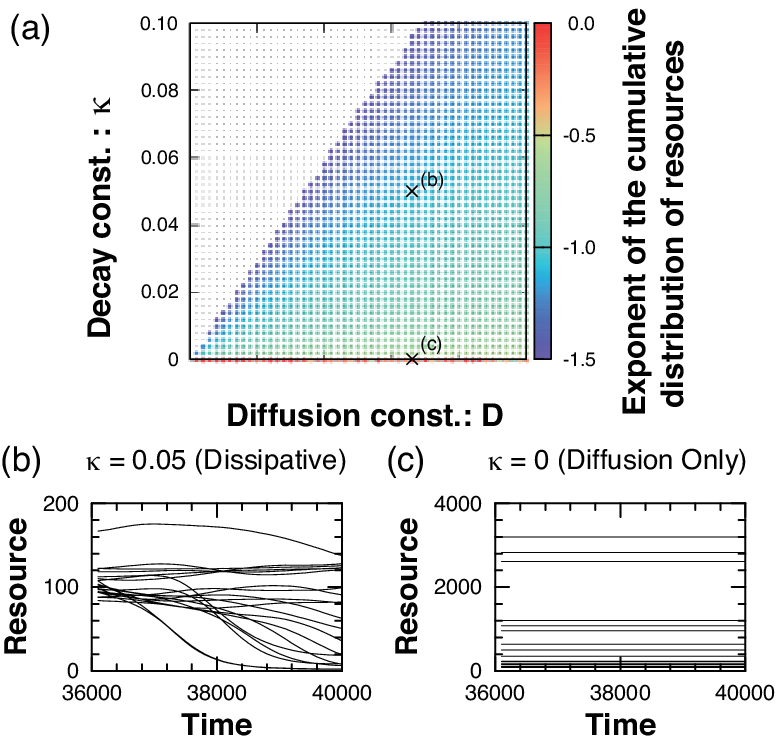}
\caption{(color).
(a)  Dependence of the exponent of the cumulative distribution of the resource on the decay constant, $\kappa$, and the diffusion constant, $D$.
The other parameter values are the same as in Fig. \ref{fig:CoEv}.
The points labeled ``(b)'' and ``(c)'' indicate the parameter values used in the corresponding cases.
(b)
Time evolution of the quantities of the resource at the 20 most resource-rich nodes in the asymptotic state
for a system with dissipation ($\kappa$ = 0.05).
In this case with nonzero $\kappa$, the distribution of the resource over the network realizes a steady state
asymptotically, but, as seen here, the quantity of the resource at each node and the rankings
of the nodes continue to change indefinitely.
(c)
Same as in (b), but with no dissipation ($\kappa$ = 0).
In this case, the quantities of the resource at the top 20 nodes remain fixed.}
\label{fig:Stationary}
\end{figure}

In the following, we focus on the extreme case $\kappa$ = 0, which can be readily treated analytically.
In this case, the resource merely diffuses over the network without dissipation, and the total quantity of the resource is conserved. Furthermore, to obtain insight into the interplay between the diffusive resource and the weight dynamics,
here we investigate the situation in which the resource dependence of the weight dynamics takes the generalized form
$$
w_{ij}(t+1)  - w_{ij}(t) =  \epsilon \left[  x_i(t)^\alpha x_j(t)^\alpha - w_{ij}(t) \right]. \eqno(3)
$$
We determine the types of organized network structure realized as a function of the parameter $\alpha$,
which controls the non-linearity of the resource dependence.
Figure \ref{fig:DiffusionOnly} summarizes the numerical results.
As seen there, two distinct types of network structure appear through the dynamics,
as determined by the value of $\alpha$.
In the case $\alpha \ge 1$, there emerges a power-law resource distribution in which
a few very resource-rich nodes (hubs) and many resource-poor nodes coexist.
Contrastingly, in the case $\alpha <1$, the resource is distributed almost evenly among all nodes,
and no prominent hubs appear.

We now analyze the asymptotic behavior of the system, making some simplifying approximations
that allow us to extract meaningful results. 
Let us first consider separately the dynamics of the resource distribution and set of link weights
in the artificial cases that the other is held fixed.
First, if the link weights were static, the quantities $x_{i}$ would converge to the equilibrium solution $x_{i}^{*} = s_{i} / (\sum_{k} s_{k}/N)$.
Second, if the resource distribution were held fixed,
the weights would relax to the solution $w_{ij}^{*} = (x_i x_j)^{\alpha}$.
With these considerations, we conjecture that the asymptotic dynamics can be approximated
by updating the variables $x_{i}$ and $w_{ij}$ in an alternating manner using two maps,
$x_{i}(n+1)=  s_{i}(n) / (\sum_{k} s_{k}(n)/N)$ and then $w_{ij}(n+2)=(x_i(n+1) x_j(n+1))^{\alpha}$,
where $n$ denotes the number of the iteration.
Although this approximation might be too crude, the results it produces exhibit reasonable agreement with the numerical simulations of the original system, as shown below.
Next, using the first of the above maps, we can eliminate $x_i(n)$ in the second.
Then, using the update rule for the strength $s_{i}$, we obtain
$$
  s_{i}(n+1)  = c_{n} s_{i}(n)^{\alpha}  \sum_{j=1}^{N}  a_{ij} s_{j}(n)^{\alpha}, \eqno (4)
$$
where $c_{n} = 1/ (\sum_{k} s_{k}(n)/N)^{2\alpha}$. 
In addition, we consider the case in which  the network topology $a_{ij}$ is chosen as a regular random graph, which means that each $a_{ij}$ is selected randomly, but $\sum_{j} a_{ij}$ is a predetermined constant, $k_{0}$.
Then, using a mean-field approximation,
we finally obtain the following relation for the strength after $n$ iterations with the initial value $s_{i}(0)$: 
$$
s_{i}(n)\propto s_{i}(0)^{\alpha^{n}}. \eqno (5)
$$
The asymptotic behavior resulting from the above maps can be classified into three cases,
corresponding to three types for the value of $\alpha$: $\alpha>1$, $\alpha<1$ and $\alpha=1$.
We now consider these individually.

In the case $\alpha >1$,
for a given initial strength distribution $P^{0}(s)$,
we have
$
  P^{n}(s) \propto P^{0} (s^{\alpha^{-n}} ) s^{-1 +\alpha^{-n}}.
$
In the asymptotic limit $n \to\infty$, the distribution $P^{n}(s)$ behaves as
$
  P^{n}(s) \to  s^{-1}.
$
In Fig. \ref{fig:DiffusionOnly}(b), it is seen that this power-law strength distribution appears
not only in the case of a regular random graph but also in the case of an ER random graph.
This suggests that the above analytic result holds for a more general network topology.
As shown in the top graph of Fig. \ref{fig:DiffusionOnly}(b),
an increase in the link weight between two nodes tends to become larger when 
the quantities of the resource at these nodes increase.
This leads to the emergence of hub nodes.
This scaling-up effect represents a kind of  ``rich get richer'' or ``economies of scale'' behavior.

In the opposite case, $\alpha < 1$, $s_{i}(n)$ in Eq. (5) converges to a uniform constant in the limit $n \to \infty$;
i.e., Eq. (5) becomes $s_{i}(n) \propto s(0)^{0}$. 
The validity of this theoretical prediction has been confirmed numerically for regular random graphs, as shown in Fig. \ref{fig:DiffusionOnly}(b).
For ER random graphs, owing to the variability of the degree, the strengths do not converge to identical values
but, rather, to some distribution with finite variance.
Note that the strength distributions for both types of random graphs in the case $\alpha <1$
possess a characteristic scale, which implies a finite mean strength.

In the critical case, $\alpha$ = 1, the situation is very delicate.
Theoretically, according to Eq (5), the strength after $n$ iterations should be given by $s(n)=s(0)$,
and thus the strength distribution should not change.
However, this prediction is inconsistent with the numerical results,
as we have already seen that a power-law strength distribution is realized for both regular random graphs and ER random graphs (see Fig. \ref{fig:DiffusionOnly}).
For complete graphs, however, the situation is different.
In this case, the strength distribution remains unchanged, as predicted by the analytic treatment (data not shown).
These results lead us to conclude that a more accurate approximation of the asymptotic dynamics is required \footnote{The direct numerical simulation of Eq (4), without the mean-field approximation, yields results consistent with those obtained from the original equations, (1) and (3).}.

In the case with no dissipation ($\kappa = 0$), the exponent of the power-law distribution is always $-1$, provided that $\alpha \ge 1$.
This type of strength distribution has been reported for the global cargo shipping network \cite{Kaluza19012010}.
In the general case with dissipation ($\kappa \neq 0$), 
the exponent of the resource power-law distribution generally depends on the parameter values,
such as the decay constant, $\kappa$, and the diffusion constant, $D$.
Figure \ref{fig:Stationary}(a) plots the exponent of the cumulative distribution of the resource in ($\kappa$,$D$) space.
These results were obtained 
by fitting the numerically generated distributions to the form $x^{\gamma}$ with a least-squares fit.
As seen there, the exponent decreases to about -1.5 with increasing $\kappa$ and decreasing $D$, and eventually
the resource distribution ceases to be of a power-law type.
We thus see that the resource disparity among the nodes is an increasing function of $D$ and a decreasing function of $\kappa$.

Interestingly, even in the case that the system has realized its asymptotic, stationary power-law distribution,
the resource ranking among  the nodes continues to change through the co-evolving dynamics.
In Fig. \ref{fig:Stationary}(b), it is seen how the quantities of  the resource possessed by the top 20 nodes
(according to the ranking at $t$ = 36000) gradually change in time from $t$ = 36000 to $t$ = 40000,
with the rankings occasionally being exchanged.
The link weights similarly continue to evolve over the entire network, even in the asymptotic regime.
This kind of dynamical phase of an organized scale-free network is observed generally in the case $\kappa \neq 0$. Note that this dynamical system described by Eqs. (1) and (2) includes no random noise.
Contrastingly, in the  case   $\kappa$ = 0, the resource ranking of the nodes becomes fixed asymptotically (see Fig. \ref{fig:Stationary}(c)).
Therefore, in addition to elucidating the statistical characteristics of the network emerging asymptotically, our model is able to describe the microscopic dynamics exhibited by a dynamical state in which both the link weights and the resource distribution continue to change indefinitely (see Fig. S2 in the Supplemental Material).
This suggests that our model may be applicable to the investigation of the vicissitudes of social phenomena,
including the dynamics of business activity, webpage access rankings and city populations \cite{Gautreau02062009}.

In conclusion, we have proposed a simple model of a co-evolving weighted network
exhibiting a dissipative diffusion process of a resource over a weighted network and resource-dependent
evolution of the network link weights.
We have demonstrated numerically and analytically that both the resource and weight distributions exhibit power-law forms in the asymptotic state as a result of the interplay between these two types of dynamics.
We believe that the most important finding of this paper is
the existence of a dynamical phase of the organized scale-free network
for a system with dissipation.
From this, we conclude that
our model can treat the dynamical formation of microscopic structure in a weighted network.
We believe that our model provides a useful basic framework for the modeling of co-evolving weighted networks
and that through the application of various generalizations,
it should be useful for investigating real-world systems of many kinds.
For example, generalizing the model to include reaction dynamics with multiple resources,
it would be readily applicable to a wide range of actual physical and social systems,
including systems driven by chemical reaction-diffusion dynamics, predator-prey dynamics and population dynamics.

We thank N. Masuda for fruitful discussions.
This work was supported by KAKENHI (24120708, 24740266, 21120002, 23115511).

%

\end{document}